\documentclass[twocolumn,showpacs,preprintnumbers,amsmath,amssymb]{revtex4}

\usepackage{graphicx}
\usepackage{dcolumn}
\usepackage{bm}

\begin{document}

\preprint{APS}

\title{Block voter model: Phase diagram and critical behavior}

\author{C.I.N. Sampaio-Filho}
\email{cesampaiof@gmail.com}
\affiliation{Departamento de Fisica, Universidade Federal de Pernambuco, 50670-901, Recife-PE, Brasil}


\author{F.G.B. Moreira}
\email{brady@df.ufpe.br}
\affiliation{Departamento de Fisica, Universidade Federal de Pernambuco, 50670-901, Recife-PE, Brasil} 
\affiliation{Departamento de Fisica Teorica e  Experimental, Universidade Federal do Rio Grande do Norte, 59072-970, Natal-RN, Brasil}



\date{\today}

\begin{abstract}
We introduce and study the block voter model with noise on two-dimensional square lattices using Monte Carlo simulations and finite-size scaling techniques. The model is defined by an outflow dynamics where a central set of $N_{PCS}$ spins, here denoted by persuasive cluster spins (PCS), tries to influence the opinion of their neighboring counterparts. We consider the collective behavior of the entire system with varying PCS size.  
When $N_{PCS}>2$, the system exhibits an order-disorder phase transition at a critical noise parameter $q_{c}$ which is a monotonically increasing function of the size of the persuasive cluster. We conclude that a larger PCS has more power of persuasion, when compared to a smaller one. It also seems that the resulting critical behavior is Ising-like independent of the range of interaction.
\end{abstract}


                              
\maketitle


\section{\label{sec:level1}Introduction}

 In non-equilibrium dynamical systems there are usually no energy functions and their time evolutions are defined by dynamical rules. These rules can be divided into two groups, namely, inflow and outflow dynamics \cite{inoutflow2006}. For inflow dynamics the center spin is influenced by its nearest neighbors. A well-known example of such dynamics, in non-equilibrium systems, is the majority-vote  model in social sciences   \cite{liggett1985, oliveira1992}. On the contrary, for outflow dynamics the information flows from the center spin (or cluster of spins) to the neighborhood. The Sznajd model \cite{sznajd2000}, which was introduced to describe opinion formation in social systems, falls into this category. It is based on the fundamental social phenomenon called social validation.

In a recent article, Castellano et al. \cite{castellano2009} discuss the state of the art of the field of social dynamics. In the particular case of binary opinion dynamics they describe several modifications and applications of the original majority-vote model \cite{liggett1985}, the Sznajd model \cite{sznajd2000}, as well as the Galam model \cite{galam2002, galam2004, stauffer2004}. 
Among the motivations supporting some opinion models, we mention  the introduction of noise through the addition of specific stochastic rules in the original dynamics and the size of the group of spins that influence the opinion of other spins. It is important to note that \textit{both} these features are relevant in the definition of the present  block voter model  and, to the best of our knowledge, have not been considered earlier. 

 In this work we perform numerical simulations on two-dimensional square lattices, with $N$ sites and periodic boundary conditions, in the relevant  $ N_{PCS} \times q$ parameter space, where the noise parameter $q$ and the persuasive cluster size $N_{PCS}$, respectively, may be regarded as the {\it  social temperature} and {\it social pressure} of the system. We use finite-size scaling theory to obtain the phase diagram in the social temperature-pressure space in the thermodynamic limit.

The present block voter model introduces long-range environment-behavior interactions in the system by considering outflow dynamics. The range of the interaction is defined by the number of spins $N_{PCS}$ inside the persuasive cluster. In principle  the size of PCS can grow until to infinity (infinite range, mean-field limit). However,  in the limit $N_{PCS} \to \infty$,  in which every spin has the same strength of interaction  with each other spin on the lattice, the system's behavior should be predict by the mean-field theory with classical critical exponents, namely, $\beta = 1/2$, $\gamma=1$, and $\nu = 1/2 $ \cite{ginzburg1960, stanley71, nielsen1977}. Here we will not consider the crossover to the mean-field  critical behavior  \cite{ginzburg1960} and restrict the simulations to finite systems taking into account the relative sizes of PCS and the whole system. Note that the ratio $\frac{N_{PSC}}{N} $ goes to zero in the thermodynamic limit.

The remainder of the paper is organized in the following way: In Sec. II we introduce the model and describe the methodology used in the simulations. In Sec. III we present a discussion of our numerical results along with a finite-size scaling analysis of the relevant quantities, and we conclude in Sec. IV.

\section{Block Voter Dynamics}
\begin{figure}
\includegraphics[width=8.5cm,height=6.5cm]{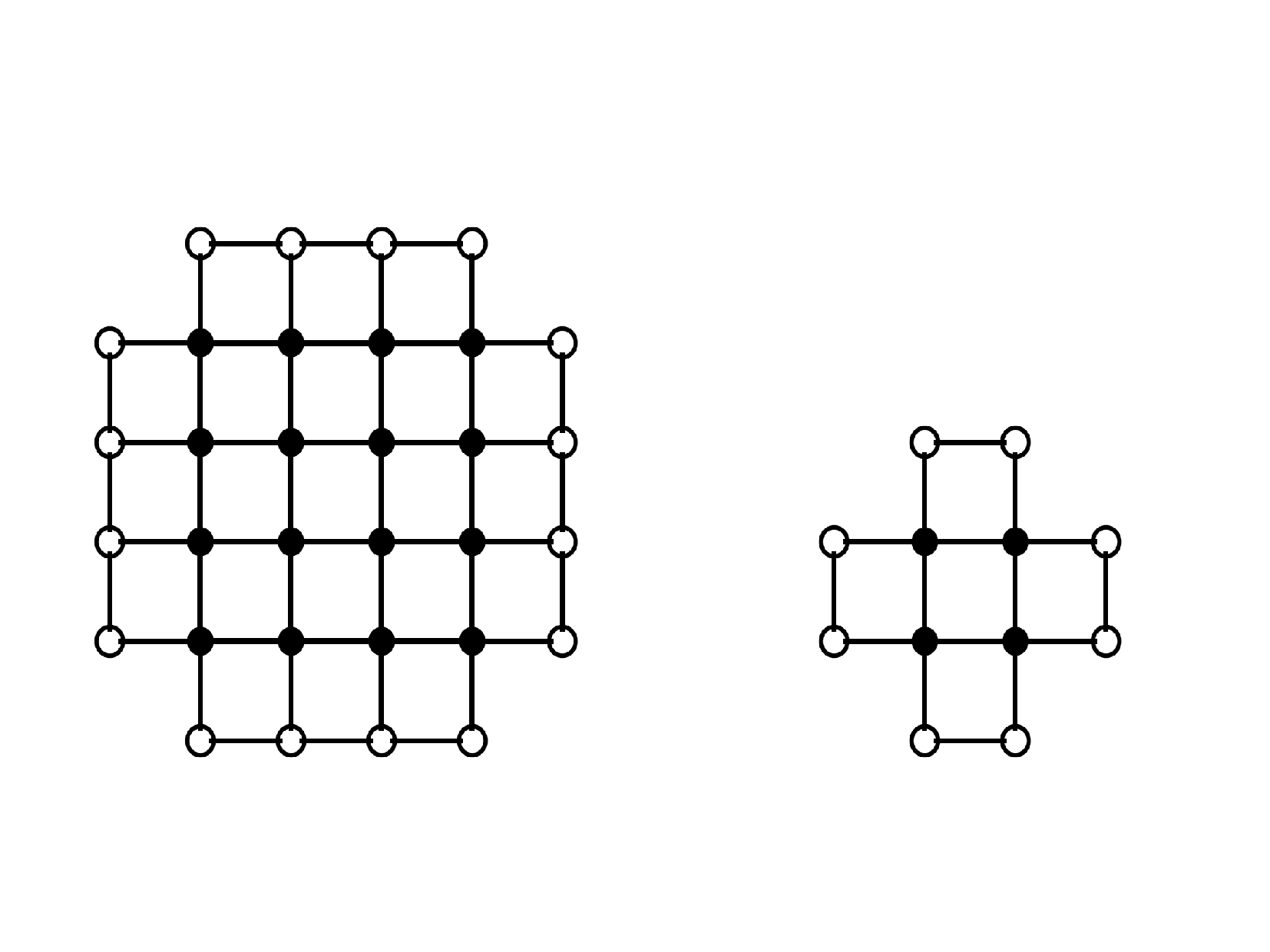}
\caption{Representative picture that explain the outflow influence, for  persuasive cluster spins (full circles) of size $N_{PCS} = 16$ and $N_{PCS} = 4$.}
\label{fig01}
\end{figure}

The model system consists of a set of $N$ two-state spin variables {$\sigma_i$}, associated with the \textit{i}th vertex of a regular square lattice of linear size $L=\sqrt{N}$. Each spin can have two possible values $\sigma_{i}=\pm 1$ corresponding to the two opposite opinions in a referendum. The system's evolution starts from a given spin configuration with periodic boundary conditions in both directions. At each time step, a block of spins randomly chosen tries to influence the spins in the neighborhood passing to them the opinion of the majority of its components. Moreover, all spins have some resistance for accepting such outflow  influence. So, independently, each spin adjacent of the $PCS$ agrees with the PCS majority state with probability ($1-q$) and  the opposite state with probability $q$. In terms of the noise parameter $q$, the rate of flipping for each adjacent spin is given by

\begin{equation}
 w(\sigma_{i})=\frac{1}{2}\left[1-(1-2q)\sigma_{i}S(\sum_{\delta = 1}^{N_{pcs}}  \sigma_{i+\delta}) \right],
 \label{eq1}
\end{equation}
where the summation is over all $N_{PCS}$ sites that make up the persuasive cluster, and $S(x) = sgn(x)$ if $x \neq 0$ and $S(0)=0$.  When $S(0)=0$, which can occur for PCS with an even number of spins, the spin $\sigma_{i}$ is flipped with probability $1/2$.  From equation (1), we notice that the block voter model shares features with the  majority-vote model \cite{tania1991, oliveira1992, brady2005}; e.g., both models exhibit up-down symmetry and  are endowed  with spontaneous broken symmetry on the parameter $q$, having then an order-disorder phase transition. In addition both non-equilibrium dynamical models do not satisfy the  condition of detailed balance, and, therefore, cannot be described by a Hamiltonian. 
However  the majority-vote model is  a kind of inflow dynamics which takes into account only the opinion of the neighboring spins of a selected node $i$. Here we consider the majority opinion of a block of spins of size $N_{PCS}$ with influence upon its neighborhood.  Fig.\ref{fig01} shows the cases of $N_{PCS} = 4$ and $N_{PCS} = 16$. 
The block voter  model is a kind of outflow dynamics defined in terms of two control parameters: $q$ and $N_{PCS}$.

In order to compare our proposal with other related dynamic rules we shall refer to  the works by de Lama et al. \cite{lama2005} and Biswas and Sen \cite{sen2009} (see \cite{castellano2009} for more detail and further literature). In \cite{lama2005} the authors consider the effect of stochastic dynamics on the Sznajd model by assuming that the original rules are fully applied  with probability $p$, while the opposite option occurs with probability $1-p$.  As $p$ is decreased the model presents a  continuous phase transition towards the stalemate state, for the case of fully connect  one-dimensional chain (mean-field limit) and for the corresponding small-world  networks. On the other hand, in \cite{sen2009}  it is proposed a two state dynamics in which the opinion of the individuals in the interface of two domains changes according to the size of their neighboring domains, which is not fixed either in time and space. The disordered version of this model, which considers a fraction of rigid spins that never change their states, the system undergoes a discontinuous phase transition from a fully ordered state (unanimity) to a disordered state where no consensus can be reached. 

The microscopic rules used in \cite{lama2005} and \cite{sen2009} are different from those of the present work in several aspects. For instance, stochastic driving is introduced in the block voter model by assuming that the majority-vote rule is applied with probability $1-q$, while the opposite option (minority) occurs with probability $q$. Moreover, we consider an outflow dynamics in which the size of the persuasive block spin is fixed in time.  Fig.\ref{fig01} illustrates the outflow influence where the central set of $N_{PCS}$ spins (full circles) tries to influence the opinion of their neighbouring counterparts (open circles). We shall nominate the central set of spins by persuasive cluster spins (PCS), and we aim in this paper to consider the collective behavior of the entire system with varying PCS size.  In short, the PCS will try to influence the neighborhood and we might expect how large it is more power of persuasion it would have.
 
To study the effect of the noise parameter $q$ and the range of interaction $N_{PCS}$ on the phase diagram and critical behaviour of the block voter model, we consider the magnetization $M_{L}$, the susceptibility $\chi_{L}$, and the Binder's fourth-order cumulant $U_{L}$ \cite{binder1981}, which are defined by:
\begin{equation}
 M_{L}(q) = \left<\left< \Bigg| \frac{1}{N}\sum_{i=1}^{N}\sigma_{i} \Bigg| \right>_{time} \right>_{sample}
 \label{eq2}
\end{equation}

\begin{equation}
 \chi_{L}(q) = N \left[\left< \left< m^{2} \right>_{time}  - \left< m \right>_{time}^{2} \right>_{sample}\right]
 \label{eq3}
\end{equation}

\begin{equation}
 U_{L}(q) = 1 - \left< \frac{\left< m^{4} \right>_{time}}{3\left< m^2 \right>_{time}} \right>_{sample}
 \label{eq4}
\end{equation}
where $N=L^{2}$ is the number of spins in the system. The symbols $<\cdots>_{time}$ and $<\cdots>_{sample}$, respectively,  denote time averages taken in the stationary state and configurational averages taken over several samples. 

For a fixed value of $N_{PCS}$, we have performed Monte Carlo simulations on regular square lattices of sizes $L=100, 120, 130, 140, 150, 160,$ and $180$. The values of $N_{PCS}=4, 9, 16,25, \dots, 100$ used satisfy the relation $\frac{N_{PSC}}{N} \le 1\%$ between the size of the persuasive cluster and the whole system.
Time is measured in Monte Carlo step (MCS), and $1$ MCS corresponds to $N$ attempts of changing the states of the spins. So, for a given $PCS$ we consider two procedures:  For  \textit{asynchronous update} we choose randomly a spin adjacent to the $PCS$ and try to flip it with   probability given by Eq. (\ref{eq1}). By repeating this procedure $N$ times, we have accomplished one MCS. In the case of \textit{synchronous update}, however, considering that for  a given $N_{PCS}$ we have $N_{adj}$ adjacent spins, we can  update simultaneously all theses spins  with the same probability [Eq. (\ref{eq1})]. One MCS is accomplished after repeating this procedure $\overline{N} = \frac{N}{N_{adj}}$ times.
For each kind of update, we waited $10^{4}$ MCS to make the system  reach the steady state, and the time averages were estimated from the next  $10^{4}$ MCS. In the critical region larger runs were needed, and we used $3 \times 10^{4}$ MCS to reach the steady state and $20 \times 10^{4}$ $MCS$ for time averages. For all set of parameters ($q, N_{PCS}$), at least $100$  independent runs (samples) were considered in the calculation of the configurational averages.  The simulations were performed using different initial spin configurations. We have checked up that  the numerical results do not depend on the initial fraction of spins in the state $\sigma=1$.

\section{Results and Discussion}

\begin{figure}
\includegraphics[width=8.5cm,height=6.5cm]{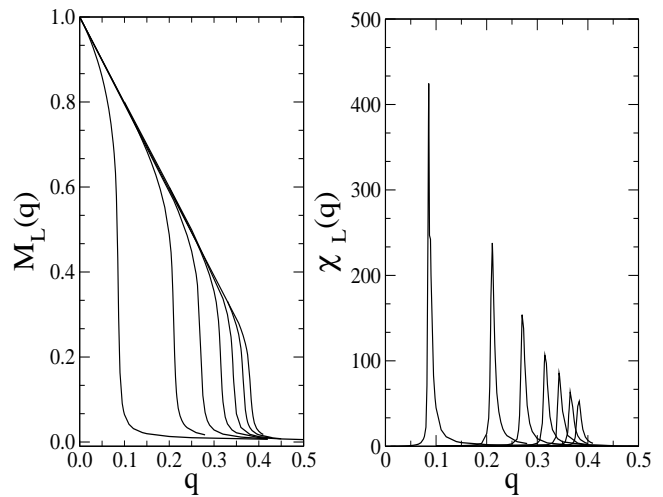}
\caption{Magnetization and susceptibility as functions of the noise parameter $q$, for $L=140$ and values of $N_{PCS}=4,9,16,25,36,49,64$ (from left to right).}
\label{fig02}
\end{figure}

Fig. \ref{fig02} shows the dependence of the order parameter $M_{L}$ and the susceptibility $\chi_{L}$ on the noise parameter, obtained from Monte Carlo simulations on square lattices with $L=140$ ($N=19600)$ and several sizes of the persuasive cluster spin, namely, $N_{PSC}= 4,9,16,25,36,49,64$. In Fig. 2(a) each curve for $M_{L}$, for a given value of $L$ and $N_{PCS}$, suggests that there exists a  phase transition from an ordered to a disordered state,  characterized by a spontaneous broken symmetry at a particular value of the noise parameter, namely, $q=q_{c}$. In the thermodynamic limit ($N \to \infty$), we should expect  that below the critical noise $q_{c}$ the system has a nonzero magnetization, whereas the magnetization vanishes for $q \ge q_{c}$. For finite systems, the value of $q$ where each curve for $\chi_{L}$ in Fig. 2(b) has a maximum is identified as $q_{c}(L)$ for the corresponding $N_{PCS}$. We note that the transition occurs at a value of the critical noise parameter which is an increasing function of the size of the persuasive cluster.

\begin{figure}
\vspace{12.4mm}
\includegraphics[width=8.5cm,height=6.5cm]{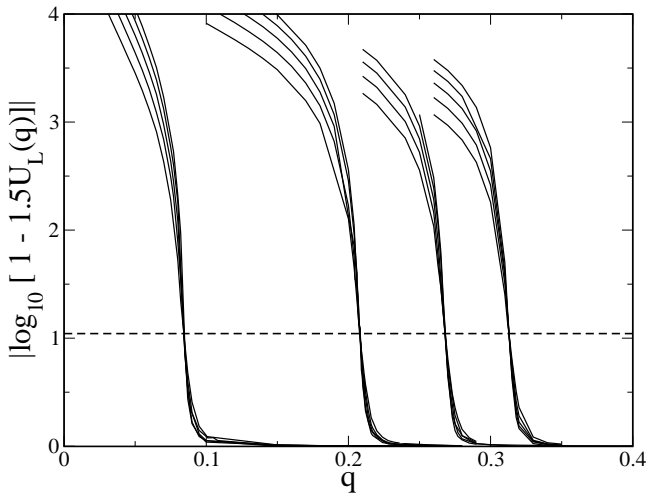}
\caption{A set of  fourth-order reduced Binder's cumulant as a function of $q$ for $N_{PCS}= 4,9,16,25$ (from left to right) and five values of lattice sizes $L$  ($L=100, 120, 140, 160$, and  $180$). The critical value $U^{*} = 0.606\pm0.004$ is shown by the horizontal line.} 
\label{fig03}
\end{figure}

For each $N_{PCS}$, we can obtain the critical value $q_{c}$ by calculating the Binder's fourth-order magnetization cumulant $U_{L}(q)$ (Eq. \ref{eq4}) as a function of the noise parameter $q$, for several lattice sizes $L$. For sufficiently large systems, these curves intercept each other in a single point $U^{*}(q_{c})$. The value of $q$ where occurs the intersection equals the critical noise $q_{c}$, which is not biased by any assumption about critical exponents since, by construction, the Binder's cumulant presents zero anomalous dimension\cite{binder1981, murilo2004}. In Fig. \ref{fig03} we plot  the reduced  fourth-order Binder's  cumulant for lattice sizes $L=100, 120, 140, 160, 180$, and four different values of $N_{PCS}$. From each set of curves,  we obtain the critical noise parameter $q_{c}$ as well as the critical value $U^{*}(q_{c})$.  As we can notice, there exists a strong dependence between the critical noise and the size of the persuasive cluster, i.e., when $N_{PCS}$ increases the critical noise also increases. 
Despite the observed dependence of $q_c$ on $N_{PCS}$, the value of the Binder's cumulant at the intersection point $U^{*}$ does not depend on the size of the persuasive cluster: We obtained $U^{*} = 0.606\pm0.004$ (discontinuous horizontal line in Fig \ref{fig03}), for all $N_{PCS}$, which is in agreement with the result for the Ising model on the regular square lattice \cite{blote1993}.
In order to  construct the phase diagram for the block voter model, we have performed this analysis for several values of the parameter $N_{PCS}$.

\begin{figure}
\includegraphics[width=8.5cm,height=6.5cm]{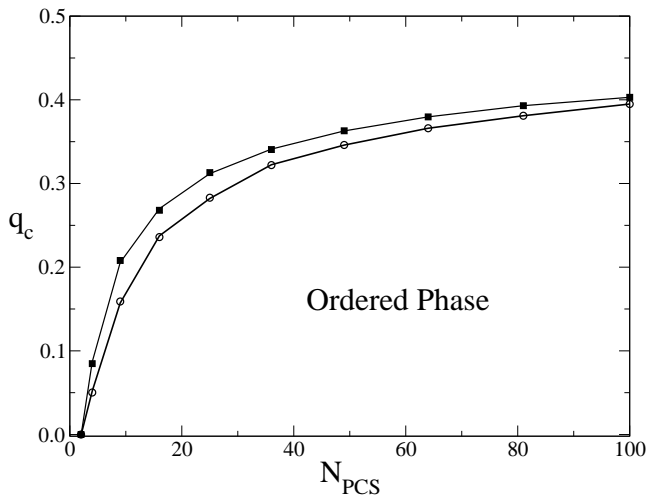}
\caption{The phase diagram of the block voter model, showing the dependence of the critical noise parameter $q_{c}$ on the size of persuasive cluster $N_{PCS}$, obtained from MC simulations using both synchronous (open symbols) and asynchronous (closed symbols) updates.}
\label{fig04}
\end{figure}


The phase diagram in the  $N_{PCS} \times q$ parameter space is shown in Fig. \ref{fig04}. The curves were obtained using  both the synchronous update with all adjacent spins to the persuasive cluster being updated at the same time, and the asynchronous one when just a single spin  randomly chosen in the neighborhood of the PCS is updated.  For both updates, there exists a phase transition only for values of $N_{PCS}>2$. Moreover, the critical noise parameter $q_c$ is an increasing function of the number of spins in the persuasive cluster. In other words, we can conclude that a larger PCS has more power of persuasion, when compared to a smaller one.

We turn now to the finite-size scaling theory \cite{fisher1972, brezin1981} that allows us to extrapolate the information available from  finite-system simulations to the relevant one in the thermodynamic limit. 
The critical behaviour of the block voter model is given by: 

\begin{equation}
 M_{L}(q) \sim L^{-\beta/\nu}\widetilde{M}(\varepsilon L^{1/\nu}),
 \label{eq5}
\end{equation}

\begin{equation}
 \chi_{L}(q) \sim L^{\gamma/\nu}\widetilde{\chi}(\varepsilon L^{1/\nu}), 
 \label{eq6}
\end{equation}

 \begin{equation}
 U_{L}(q) \sim \widetilde{U}(\varepsilon L^{1/\nu}),
 \label{eq7}
\end{equation}
where $\varepsilon=(q-q_{c})$ is the distance from the critical noise. The exponents $\beta/\nu$, $\gamma/\nu$, and $\nu$ are, respectively, associated  to the decay of the order parameter $M_{L}(q)$,  the divergence of the susceptibility $\chi_{L}(q)$, and of the correlation length ($\xi \sim \varepsilon^{-\nu})$. 
The universal scaling functions $\widetilde{M}(\varepsilon L^{1/\nu})$, $\widetilde{\chi}(\varepsilon L^{1/\nu})$, and $\widetilde{U}(\varepsilon L^{1/\nu})$ depend only on the scaling variable $x=\varepsilon L^{1/\nu}$.

The correlation length exponent $\nu$ can be obtained from the derivative of the Binder's fourth-order cumulant with respect to the noise parameter, namely, $U_{L}^{'}(q_{c})$.
Moreover, we can check our estimates for the critical exponents considering the hyperscaling relation derived  from the  Rushbrooke and Josephson scaling laws, namely:
\begin{equation}
 2\beta/\nu + \gamma/\nu =  d,
 \label{eq8}
\end{equation}
which is valid for Euclidean dimension $d$ less than the upper critical dimension $d_{u}$ \cite{upperDim2008}.

\begin{table}
\vspace{3mm}
\caption{\label{tab:table1} Results for the critical noise $q_{c}$, the critical exponents $\beta/\nu, \gamma/\nu, 1/\nu$ and the effective dimensionality $D_{eff}$ for the Block Voter Model on regular square lattices considering different sizes of persuasive cluster spin $N_{PCS}$.}
\begin{ruledtabular}
\begin{tabular}{cccccc}
 $N_{PCS}$ &$q_{c}$  &$\beta/\nu$ &$\gamma/\nu$ &$1/\nu$ & $D_{eff}$ \footnotemark[1] \\
\hline
 $4$    &$0.0846(2)$       & $0.130(2)$          &$1.72(4)$               & $1.05(2)$                         &$1.98(4)$ \\
 9        &$0.2080(1)$       &$0.124(4)$           &$1.76(3)$               & $0.98(1)$                         & $2.01(4)$ \\
 16      &$0.2680(1)$       &$0.114(2)$           &$1.73(1)$               & $0.96(2)$                         &$1.95(2)$ \\
 25      &$0.3130(4)$       &$0.132(1)$           &$1.74(1)$               &$0.95(2)$                          &$2.00(1)$  \\
 36      &$0.3406(3)$       &$0.130(3)$           &$1.70(3)$               &$0.96(3)$                          &$1.96(4)$   \\
 49      &$0.3630(1)$       &$0.113(3)$           &$1.70(2)$               &$0.96(2)$                          &$1.96(3)$ \\
 64      &$0.3790(2)$      &$0.118(2)$            &$1.71(4)$               &$0.96(3)$                          &$1.95(4)$\\
\end{tabular}
\end{ruledtabular}
\footnotetext[1]{Obtained using $D_{eff}=2\beta/\nu + \gamma/\nu$}
\end{table}


We have explored the dependence of the magnetization and susceptibility on system size $L$, at $q=q_{c}$, when different values of $N_{PCS}$ are considered. From the scaling relations (\ref{eq5}) and (\ref{eq6}), respectively, for a given $N_{PCS}$ the corresponding straight lines in these log-log plots have slopes equal to the exponents $\beta/ \nu$ and $\gamma/ \nu$.  This analysis yields exponents very close to the two-dimensional Ising exponents, namely, $\beta/\nu=0.125$ and $\gamma / \nu = 1.75$. Similarly, from the slopes of the resulting straight lines in the log-log plots of the  derivative of the Binder's cumulant with respect to the noise parameter, at $q=q_{c}$, we determine the exponent associate with the correlation length in good agreement with the exact value $\nu=1.0$. Finally, we have checked whether the calculated exponents satisfy the hyperscaling relation (\ref{eq8}) with $D_{eff}=2\beta/\nu + \gamma/\nu$. For all PCS sizes considered we obtained $D_{eff}=2$, within error bars. Our results are summarized in Table I.

According to Grinstein et al's conjecture \cite{grinstein1985}, all equilibrium models with up-down symmetry on regular lattices have Ising-like critical behaviour. This conjecture has also been verified for several nonequilibrium model systems \cite{costa2005, bennett1985, wang1988, marques1990}.  Moreover, by considering the universality of the critical exponents for arbitrarily large but finite range of the interactions (see, e.g.,  \cite{MonAndBinder1993} and references therein), we actually should not expect even such a small dependence in the calculated exponents with the number of persuasive spins $N_{PCS}$.  In fact, the present non-equilibrium  Block Voter model and the equilibrium (finite-range interaction) Ising model are described by the same critical exponents.

\begin{figure}
\vspace{8mm}
\includegraphics[width=8.5cm,height=6.5cm]{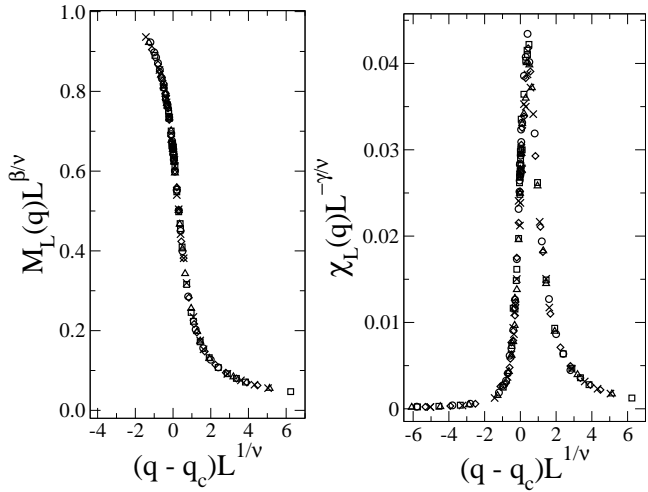}
\caption{Data collapsing of the order parameter and of the susceptibility  for $N_{PCS}=9$, with five different values of $L=100, 120, 140, 160,$ and $180$. The universal functions are consistent with Ising exponents: $\beta/\nu = 0.125$, $\gamma/\nu = 1.75$ and $1/\nu = 1.0$.}
\label{fig07}
\end{figure}

In order to quantify the above statement,  we shall consider a more accurate analysis that consists of obtaining the universal curves for the magnetization $\widetilde{M}(x) = M_{L}(q)L^{\beta/\nu}$ and for the susceptibility $\widetilde{\chi}(x) = \chi_{L}(q)L^{-\gamma/\nu}$ with $x = (q-q_{c})L^{1/\nu}$, which follows  from Eqs. (5) and (6), respectively, and represent the data collapsing for fixed $N_{PCS}$ and different sizes $L$ of the system. It is worth to note that the universal curves are obtained once the correct values for the exponents are used. Fig. \ref{fig07} shows the data collapse for the case of $N_{PCS}=9$ and five different system sizes $L=100, 120, 140, 160,$ and $180$, using the following exponents: $\beta/\nu = 0.125$, $\gamma/\nu = 1.75$ and $1/\nu = 1.0$. A similar analysis applied to other values of $N_{PCS}$ supports the conclusion that the present non-equilibrium  model is in the same universality class of the two-dimensional equilibrium Ising model.

The relaxation time  scales with system size as $\tau\sim L^{z}$, where $z$ is the dynamical critical exponent. Moreover,  we have the finite-size scaling relation for the second moment \cite{Okano1997}
\begin{equation}
 M^{(2)}(t) \sim t^{(d-2\beta/\nu)/z}.
 \label{eq9}
\end{equation}
Using $d=2$ and $\beta/\nu=0.125$, we obtained  $z=2.20(3)$ from short-time simulations of the block voter model with different values of $N_{PCS}$. This value  should be compared with z=2.137(11) and z=2.155(03) for the Ising model using Metropolis and heat-bath dynamics \cite{Okano1997}.

We conclude this Section with some comments on the equivalence between the inflow and outflow versions of the block voter dynamics, as should be expect since the direction of the flux of information does not affect the symmetry of the model. In order to show this equivalence, we simulated the inflow version of our model for some sizes of  the persuasive cluster of spins. Besides the Ising critical behavior, we obtained the same dependence of the critical parameter $q_c$ on $N_{PCS}$ shown in Fig. 4.

\section{Conclusion}

In this paper, Monte Carlo simulations and finite-size scaling theory was used to study a medium-ranged interactions version of voter model with outflow dynamics. The model is defined in terms of two parameters, the size of the persuasive cluster of spins ($N_{PCS}$) and the noise parameter $q$ associated with the resistance that every spin has for accepting such outflow influence. Considering  both synchronous and asynchronous updates, the resulting  phase diagram in the $ N_{PCS}\times q$ parameter space indicates that the region where there exists an ordered phase increases with increasing range of the interactions, meaning that a larger PCS has more power of persuasion, when compared to a smaller one. The calculated critical exponents for different sizes of PCS support the well-known criterion of universality class, stating that medium-range interactions models with up-down symmetry exhibit Ising-like critical behavior for all arbitrarily large but finite range of the interactions.

\begin{acknowledgments}
C.I.N. Sampaio-Filho is supported by CNPq. The authors acknowledge partial support from CNPq, FINEP and FACEPE. 
\end{acknowledgments}


\bibliography{blockVoterRevised}

\end{document}